\documentclass[11pt]{article} 

\title{
{ Bounds for randomly shared risk of heavy-tailed loss factors}\\
}

\author{Oliver Kley\thanks{Center for Mathematical Sciences, Technische Universit\"at M\"unchen,  85748 Garching, Boltzmannstrasse 3, Germany, e-mail: oliver.kley@tum.de\,,\,cklu@tum.de}
\and 
Claudia Kl\"uppelberg\footnotemark[1],
}
\usepackage{amssymb}
\usepackage{amsfonts}
\usepackage{amsmath}
\usepackage{amsthm}
\usepackage{graphicx}
\usepackage{caption}
\usepackage{subcaption}
\usepackage[usenames, dvipsnames]{xcolor}
\usepackage{verbatim}
\usepackage{dsfont}
\usepackage{color}
\usepackage[numbers]{natbib}
\usepackage{relsize}
\usepackage{lmodern}
\usepackage{url}
\usepackage{siunitx}
\usepackage{MnSymbol}
\usepackage[english]{babel}
\usepackage{tikz}
\usetikzlibrary{arrows}

\numberwithin{equation}{section}
\usepackage{caption} 
\newtheorem{theorem}{Theorem}[section]
\newtheorem{lemma}[theorem]{Lemma}
\newtheorem{remark}[theorem]{Remark}
\newtheorem{example}[theorem]{Example}
\newtheorem{proposition}[theorem]{Proposition}
\newtheorem{definition}[theorem]{Definition}

\newtheorem{corollary}[theorem]{Corollary}
\newtheorem{fig}[theorem]{Figure}

\newcommand{\bthe}{\begin{theorem}}
\newcommand{\ethe}{\end{theorem}}

\newcommand{\ben}{\begin{enumerate}}
\newcommand{\een}{\end{enumerate}}

\newcommand{\bit}{\begin{itemize}}
\newcommand{\eit}{\end{itemize}}

\newcommand{\beq}{\begin{equation}}
\newcommand{\eeq}{\end{equation}}

\newcommand{\ble}{\begin{lemma}}
\newcommand{\ele}{\end{lemma}}

\newcommand{\bde}{\begin{definition}\rm}
\newcommand{\ede}{\halmos\end{definition}}

\newcommand{\bco}{\begin{corollary}}
\newcommand{\eco}{\end{corollary}}

\newcommand{\bpr}{\begin{proposition}}
\newcommand{\epr}{\end{proposition}}

\newcommand{\brem}{\begin{remark}\rm}
\newcommand{\erem}{\end{remark}}

\newcommand{\bproof}{\begin{proof}}
\newcommand{\eproof}{\end{proof}}

\newcommand{\bexam}{\begin{example}\rm}
\newcommand{\eexam}{\end{example}}

\newcommand{\bfi}{\begin{fig}}
\newcommand{\efi}{\end{fig}}

\newcommand{\btab}{\begin{tab}}
\newcommand{\etab}{\end{tab}}

\newcommand{\beao}{\begin{eqnarray*}}
\newcommand{\eeao}{\end{eqnarray*}\noindent}

\newcommand{\beam}{\begin{eqnarray}}
\newcommand{\eeam}{\end{eqnarray}\noindent}

\newcommand{\barr}{\begin{array}}
\newcommand{\earr}{\end{array}}

\newcommand{\bdis}{\begin{displaymath}}
\newcommand{\edis}{\end{displaymath}\noindent}

\def\P{{\mathbb P}}
\def\E{{\mathbb E}}
\def\R{{\mathbb R}}

\def\P{\mathbb{P}}

\def\cals_+{{\cals_+}}

\def\cals{{\mathcal{S}}}

\def\bone{{\mathds 1}}
\def\1{\mathds{1}}

\newcommand{\stv}{\stackrel{v}{\rightarrow}}

\newcommand{\tto}{t\to\infty}
\newcommand{\xto}{x\to\infty}

\newcommand{\al}{{\alpha}}

\newcommand{\var}{{\rm Var}}

\newcommand{\ind}{{\rm ind}}
\newcommand{\dep}{{\rm dep}}

\newcommand{\halmos}{\quad\hfill\mbox{$\Box$}}  

\def\P{{\bf {\mathbb{P}}}}

\newcommand{\pr}[1]{\P\left[#1\right]}

\newcommand{\VaR}{{\rm VaR}}
\newcommand{\CoTE}{{\rm CoTE}}

\allowdisplaybreaks[4]

\begin{document}


\maketitle

\begin{abstract}
For a risk vector $V$, whose components are shared among agents by some random mechanism, we obtain asymptotic lower and upper bounds for the {individual} agents' exposure risk and the {aggregated} risk in the market.
Risk is measured by Value-at-Risk or Conditional Tail Expectation.
We assume Pareto tails for the components of $V$ and arbitrary dependence structure in a multivariate regular variation setting.
Upper and lower bounds are given by asymptotically independent and fully dependent components of $V$ {with respect} to the tail index $\al$ being smaller or larger than 1. 
Counterexamples, {where for non-linear aggregation functions no bounds are available,} complete the picture.
\end{abstract}





\section{Introduction}\label{s1}

Let $V_{j}$ for $j=1,\dots,d$ be risk variables having Pareto-tails, so that, for possibly different $K_j>0$ and tail index $\al > 0$,
\beam\label{pareto}
P(V_j>t)\sim K_j t^{-\al},\quad t \to \infty.
\eeam
(For two functions $f$ and $g$ we write $f(t)\sim g(t)$ as $\tto$ if $\lim_{\tto} f(t)/g(t) = 1$.)
We summarize all risk variables in a vector $V=(V_1,\dots,V_d)^\top$. 
{The tail index $\alpha$ is assumed to be the same for all $j=1,\dots, d$ since, when aggregating risk factors with different tail indexes, always the smallest $\alpha$ wins as a famous result of \cite{Breiman} states. 
That said, assuming the same $\alpha$ practically means to choose the subset with the smallest $\alpha$ out of a set of risk factors resulting in a dimension reduction. }

The $d$ risks in $V$ are shared among $q$ agents by some random mechanism.
Let $F_i$ denote the {\em exposure of agent $i$} and $F=(F_1,\dots,F_q)^\top$ the exposure vector. 
The risk sharing is governed by a random $q \times d$  matrix  
 $A=(A_{ij})_{i,j=1}^{q,d}$ (independent of $V$) in such a way that $F_{i}=\sum_{j=1}^{d}A_{ij}V_j$ for $i=1,\dots,q$ or, eqivalently, in matrix notation
\begin{gather}\label{F}
F=AV.
\end{gather}
{
Whether $A$ is deterministic or stochastic may depend on the quality of available information. An internal analyst or regulator with sufficient knowledge may consider $A$ as deterministic, whereas an external analyst (working for an institutional investor for instance) may consider it random due to lack of insight.}
This note is motivated by \cite{KKR2} and \cite{KKR1}, where the risk variables $V_j$ model large insurance claims and agents represent reinsurance companies. 
The claims {can for instance be} shared  randomly with a mechanism given by a bipartite graph structure, resulting in
\begin{gather}\label{eq2.2}
A_{ij} = \frac{\1 (i\sim j)}{\deg(j)},
\end{gather}
where $\1 (i\sim j)$ indicates whether agent $i$ takes a (proportional) share of risk $j$ or not, {and $\deg(j)$ denotes the total number of agents who have chosen to insure risk $j$}.
Further examples include operational risk, modelling event types (risk variables) and business lines (agents), where Pareto tails are natural (cf. \cite{OpRiskBK}), and also overlapping portfolios (common asset holding) as described in \cite{Cacciolietal}.

In all these applications it is of interest to quantify  not only the risk of single agents, but also the market risk---saying that we  mean the aggregated risk in the market---which is of high relevance to the regulator. {
Following ideas in \cite{axiomsystemic} we assess the market risk by a risk measure on the $r$-norm ($r\geq 1$) or $r-$quasinorm ($0<r<1$) $\|F\|:=\|F \|_r=\left(\sum_{i=1}^{q}F_i^r\right)^{1/r}$ of the exposure vector $F$. These aggregation functions satisfy most of the required axioms in \cite{axiomsystemic} and are continuous as well as convex or concave, respectively. 
 Our market risk measures do not necessarily satisfy the normalization condition required there:
Assuming a total unit loss split to equal parts among the agents, we have  $\|(1/q,\ldots,1/q)\|_r=q^{1/r-1}<1$ if $r>1$ and  $\|(1/q,\ldots,1/q)\|_r=q^{1/r-1} > 1 $ for $0<r<1$, hence, the normalization condition of \cite{axiomsystemic} is satisfied only if $r=1$. 
In the first case we see that the $r-$norm underestimates the additive risk by convexity. In particular, if the number of agents increases, the loss measured by the $r-$norm decreases. As a consequence, norms with $r>1$ are not suitable for systemic risk assessment as they imply hereby the possibility of regulatory arbitrage.  
We argue, however, that this underestimation may be realistic in some applications as a larger market may be less risky due to a balance of risk as is well-known for insurance portfolios. Moreover, norms of that type can be useful in portfolio analysis, see \cite{BurgertRü}.
In the second case $0<r<1$ the $r-$quasinorm will overestimate the additive risk. This situation in turn may be realistic whenever amplification mechanisms come into play as it happens with systemic risk. In addition, 
\cite{fernholz} employ $r-$quasinorms for portfolio construction. 
Our framework allows for great variability concerning the choice of the aggregation function: convex, linear or concave. The decision which aggregation function to employ is in the end application-driven and mostly a decision based on economic reasoning.
}

We investigate risk based on the Value-at-Risk ($\VaR$) and Conditional Tail Expectation ($\CoTE$), which we assess by asymptotic approximations.

Let $V_{\ind},V, V_{\dep}$ be risk vectors as above with different dependence structures among the risk variables. 
Here $V_{\ind}$ corresponds to asymptotically independent variables and $V_{\dep}$ to asymptotically fully dependent variables
in the framework of multivariate regular variation as in \cite{KKR1}.

 As in the copula world (see \cite{BRV,EPR}) it is possible to assess the two extreme dependence structures; i.e., $V_{\ind}, V_{\dep}$, and it is of high relevance to understand, if or under which conditions these extreme dependences lead to upper and lower bounds of risk for arbitrary dependence structures. Remarkably, \cite{EPR} show that the comonotonic copula does not lead to an upper bound, and a procedure is provided there to find the best possible upper VaR bound. In \cite{BRV}, information on variance is added. This reduces the set of feasible copulas. Then the upper bound can be lower than the comonotic VaR. 
The related problem of sub-  and super-addtivity  has been  investigated in \cite{Embrechts}.  We extend the setting and scope in \cite{Embrechts} significantly: first, by allowing for diversity in the tails as in \eqref{pareto} and, second, by incorporating a stochastic market structure as in \eqref{F} allowing for risk assesment in a much wider way. Moreover, the results in  
\cite{Embrechts} are also formulated for general aggregation functions but the effect of non-linearity---hence the possible breakdown of general bounds---is not considered there. In that sense, our results add  new important aspects to the existing literature.

 This note is organised as follows. In Section~2 we present $V$ as a regularly varying vector with different dependence structures.
 Here we also define the risk measures $\VaR$ and $\CoTE$ for arbitrary random variables, and summarize their asymptotic behaviour in our framework. In Section~3 we derive bounds for single agent and market risk based on asymptotically independent and fully dependent random variables. We also give counter examples to present the limitations of the bounds.

\section{Preliminaries}\label{s2}

\subsection{Multivariate regular variation}

We recall from \cite{Resnick2007}, Ch. 6 that
the positive random vector $V\in\R^d_+$  is  {\em multivariate regularly varying} if  there is a Radon measure $\nu\not\equiv 0$ on the Borel $\sigma$-algebra $\mathcal{B}= \mathcal{B}(\R_+^d \setminus \{0\})$, where $0$ denotes the zero vector in $\R^d$,   such that
\begin{gather}\label{defMVR}
n\pr{{n^{-1/\al}V}\in \cdot}\stackrel{v}{\rightarrow} \nu(\cdot),\quad n \rightarrow \infty.
\end{gather} 
The symbol $\stv$ stands for vague convergence.
Moreover, the measure $\nu$ is homogeneous of some  order $-\alpha$ with $\alpha>0$ and  is called the {\em exponent measure of $V$}.

{ We fix a norm $\| \cdot\|$ on $\R^d $ in such a way that for all canonical unit vectors $\|e_j\|=1,\ j=1,\dots,d.$ 
This actually entails a slight abuse of notation as we also write $\|\cdot\|$ for the aggregation function of the vector $F$ of agent exposures on $\R^q$}. 
Denoting by $\mathbb{S}_{+}^{d-1}=\{x \in \R^{d}_{+}:\ \|x\|=1 \}$ the positive sphere in $\R^d$ , the existence of the exponent measure $\nu$ is equivalent to the existence of a Radon measure $\rho\not\equiv 0$ on the Borel $\sigma$-algebra $\mathcal{B}(\mathbb{S}_{+}^{d-1})$ in such a way that for all $u>0$ 
\begin{gather}\label{spectraldef}
\frac{\pr{\|V \|>ut, V\|V\|^{-1}\in \cdot}  }{\pr{\|V \|>t }} \stv u^{-\al}\rho(\cdot),\quad t \rightarrow \infty, 
\end{gather}
holds. The measure $\rho$ is called the {\em spectral measure of $V$}.
The precise relation between $\nu$ and $\rho$ can be found in \cite{Resnick2007},  Ch.~6.

Finally, we note that convergence in \eqref{defMVR} also implies 
\begin{gather}\label{insteadmu}
\frac{\pr{ t^{-1}V \in \cdot  } }{\pr{\|V\|>t}  } \stv \frac{\nu{(\cdot)}}{\nu(\{x: \|x\|>1\})},\quad t \rightarrow \infty.
\end{gather}
The {\em  tail index} $\al > 0$ is also called the  index of regular variation of $V$, and we write $V \in \mathcal{R}(-\alpha).$


We shall often work with the so-called  {\em  canonical exponent measure} $\nu^*$ of $V$, which is defined as  the image measure $\nu^{*}=\nu\circ T$ under the  transformation mapping $T:\R_+^{d}\rightarrow \R_+^{d}$, given by 
$$T(x)=(\nu(\{ x_1>1\})^{1/\al}x_1^{1/\al},\dots,\nu(\{ x_d>1\})^{1/\al}x_d^{1/\al}  )^{\top}.$$
Then $\nu^*$ has standardized margins and a tail index $1$, corresponding to $P(V_j>x) \sim x^{-1}$ as $\xto$. 

The corresponding spectral measure $\rho^{*}$ is called the   {\em  canonical spectral measure} and is characterized by 
\begin{gather}\label{canspec}
\int_{\mathbb{S}_{+}^{d-1}} s_j \rho^{*}(ds)=1,\quad  j=1,\dots, d,
\end{gather}
see \cite{Beirlant}, p.~259. 

For the matrix $A$ and a given norm $\|\cdot\|$, which gives rise to an operator norm 
$$\displaystyle\|A\|_{{\rm op}}=\sup_{\|x\|=1} \|Ax\|,$$
we require througout the following:
\begin{itemize}
\item
  $A$ satisfies the moment condition $\E \|A\|_{\operatorname{op}}^{\alpha + \delta}<\infty $ for some $\delta>0$  and $\al$ as in \eqref{pareto};
\item
the vector $V$ is independent of the random matrix $A$, while $V_1,\ldots,V_d$ may not be independent of each other.
\end{itemize}
{If both conditions hold, then the vector $F=AV$  is again regularly varying with exponent measure $\E\nu\circ A^{-1}$} (cf. \cite{Basrakarticle}, Proposition A.1).
\subsection{Risk measures}

We also recall the following risk measures.

\bde\label{defVaRCoTE}
The {\em Value-at-Risk (\VaR)} of a random variable $X$ at confidence level $1-\gamma$ is defined as
\begin{gather*}
\VaR_{1-\gamma}(X):=\inf\{ t \geq 0: \pr{ X>t } \leq \gamma  \},\quad \gamma \in (0,1),
\end{gather*} 
 and the {\em Conditional Tail Expectation (\CoTE)} at confidence level $1-\gamma$, based on the corresponding \VaR,  as
\begin{gather*}
\CoTE_{1-\gamma}(X):=\E[X \mid X> \VaR_{1-\gamma}(X)],\quad \gamma \in (0,1).
\end{gather*}
\ede

Throughout the following constants will be relevant
\beam\label{constants}
C^i_{\ind}=\sum_{j=1}^d K_j \E  A_{ij}^\al,\quad i=1,\ldots,q,\quad & \mbox{and} & \quad C^S_{\ind}=\sum_{j=1}^d K_j\E\|A e_j\|^\al, \\ 
C^i_{\dep} =\E(AK^{1/\al}\bone)^\al_i,\quad i=1,\ldots,q,\quad & \mbox{and} & \quad C^S_{\dep}=\E\|AK^{1/\alpha}\bone\|^\al,
\eeam
{where we summarize the constants $K_j$ for $ j=1,\dots, d$ from \eqref{pareto} in a diagonal matrix 
\begin{gather}\label{defK} K^{1/\al}:=\operatorname{diag}(K^{1/\al}_1,\dots,K_d^{1/\al}).\end{gather} }
\ble[\cite{KKR1}, Corollaries~3.7 and~3.8]\label{varandes}
Let  $F=AV = (F_{1},\dots,F_{q})^{\top}$.\\
(a) \, {\em Individual risk measures:}\\
For $\al>0$ the individual Value--at--Risk of agent $i\in\{1,\ldots,q\}$ {satisfies} 
 \begin{gather}\label{uniVaRasym}
\VaR_{1-\gamma}(F_{i})\sim C^{1/\alpha} \gamma^{-1/\alpha} ,\quad \gamma\rightarrow 0.
\end{gather} 
For $\al>1$ the  individual Conditional Tail Expectation of  agent $i\in\{1,\ldots,n\}$ {satisfies}
$$\CoTE_{1-\gamma}(F_{i}) \sim \frac{\alpha}{\alpha-1} \VaR_{1-\gamma}(F_{i})\sim \frac{\alpha}{\alpha-1} C^{1/\alpha} \gamma^{-1/\alpha}   \,,\quad \gamma\rightarrow 0.$$ 
The individual constants are either $C= C^i_{\ind}$ or $C=C^i_{\dep}$ for  $V_1,\dots,V_d$ asymptotically independent or asymptotically fully dependent, respectively.\\
(b) \, {\em Market risk measures:}\\
The market Value--at--Risk of the aggregated vector $\|F\|$ satisfies
\begin{gather}\label{marketVaRasym}
\VaR_{1-\gamma}(\|F\|) \sim C^{1/ \alpha}\gamma^{-1/\alpha},\quad \gamma\rightarrow 0.
\end{gather}
If $\alpha >1$ the market Conditional Tail Expectation of  the aggregated vector $\|F\|$ satisfies
$$\CoTE_{1-\gamma}(\|F\|) \sim \frac{\alpha}{\alpha-1} \VaR_{1-\gamma}(\|F\|)\sim \frac{\alpha}{\alpha-1} C^{1/ \alpha}\gamma^{-1/\alpha}  \,,\quad \gamma\rightarrow 0.$$ 
 The market constants referring to the system setting are either $C= C^S_{\ind}$ or $C=C^S_{\dep}$ for  $V_1,\dots,V_d$ asymptotically independent or asymptotically fully dependent, respectively.
\ele

\section{Bounds for general dependence structure}\label{s3}
 
Recall from {(3.12) and (3.14)} of \cite{KKR1} that the constants \eqref{constants} can be expressed in terms of the exponent measure via
\begin{gather}
\label{constantsnuind} C^i_{\ind}=\E\nu_{\ind}\circ A^{-1}(\{x: x_i>1 \}),\ i=1,\ldots,q,\quad\mbox{and}\quad C^S_{\ind}=\E\nu_{\ind}\circ A^{-1}(\{x: \|x\|>1 \}) \\ 
\label{constantsnudep}C^i_{\dep} =\E\nu_{\dep}\circ A^{-1}(\{x:x_i>1  \}),\ i=1,\ldots,q,\quad\mbox{and}\quad C^S_{\dep}=\E\nu_{\dep}\circ A^{-1}(\{x:\|x\|>1|  \})
\end{gather}
with (cf. Lemma~2.2 of  \cite{KKR1})
\begin{gather}
\nu_{\ind}([0,x]^{c})=  \sum_{j=1}^{d} K_j x_{j}^{-\alpha}\ \ \text{ and}\ \ \nu_{\dep}([0,x]^{c})= \max_{j=1,\dots,{d}} \{K_j x_{j}^{-\alpha}\}.
\end{gather}
The analogues of the constants $C^{i}_{\ind}$, $C^{i}_{\dep}$ as well as $C^{S}_{\ind}$ and $C^S_{\dep}$ in the case of an arbitrary extremal dependence structure of the vector $V$, represented by some exponent measure $\nu$ with $\nu_{\ind}\neq\nu\neq\nu_{\dep}$, are then 
\begin{gather}
C_{\nu}^i= \E\nu\circ A^{-1}(\{ x: x_i>t\})\ \ \text{and} \ \ C_{\nu}^S= \E\nu\circ A^{-1}(\{ x: \|x\|>t\}). 
\end{gather}
{In the light of Lemma~\ref{varandes} it suffices to determine bounds for the constants $C_{\nu}^i$ and $C_{\nu}^{S}$   in order to obtain  asymptotic bounds for $\VaR$ or $\CoTE$ in the respective cases.} 

With $K^{1/\alpha}$ from \eqref{defK},  for the exponent measure $\nu$ of the vector $V$ with any dependence structure, we obtain 
$$C_{\nu}^{S}=\E\nu\circ K^{1/\al}\circ(A K^{1/\al} )^{-1}(\{ \|x\|>1 \}) \quad\mbox{and}\quad C_{\nu}^{i}=\E\nu\circ K^{1/\al}\circ(A K^{1/\al} )^{-1}(\{ x_i>1 \}).$$ 
Note that the measure $\nu\circ K^{1/\al}$ has balanced tails; i.e.,  $\nu\circ K^{1/\al}(\{ x_j >1 \})=1, j=1,\dots,d.$
Since all marginal random variables are as in \eqref{pareto}, regardless of the dependence structure of the vector $V$,
for the proofs of all theorems below we can and do assume that margins are standardized; e.g. $K_j=1$ for $j=1,\dots,d$.  
Moreover, for establishing inequalities between $C^{i}_{\ind}, C^{i}_{\dep}$ and $C^{i}_{\nu}$ or $C^{S}_{\ind}, C^{S}_{\dep}$ and $C^{S}_{\nu}$, respectively, it is sufficient to prove the corresponding inequalities for all realizations of the random matrix $A$. 
We obtain the following bounds for the constants defining the individual risk measures.

\bthe\label{theindiv}
Let the three $d$-dimensional vectors  $V_{\ind}$, $V$ and $V_{\dep}$ be given with equal margins $V_1,\dots,V_d$ with $\pr{V_j>t}\sim K_j t^{-\al}$, {but different exponent measures $\nu_{\ind}, \nu, \nu_{\dep}$}.
Then for the  constants $C^{i}$ referring to agent $i$ the  following inequalities hold:
\beam
\label{Calphagreateroneindiv}C^{i}_{\ind} \, \leq \, C_\nu^{i} \, \leq \, C^i_{\dep} \quad & \operatorname{for } \alpha\geq 1,\\
\label{Calphasmalleroneindiv}C^{i}_{\dep}  \, \leq \, C_\nu^{i} \, \leq C^i_{\ind} \, \quad & \operatorname{for } \alpha< 1.
\eeam
\ethe

\begin{proof}
Let $a_i:=A_{i\cdot}$ be the $i$-th row of the matrix $A$ and $V_{\ind},V, V_{\dep}$ be as above the risk vectors with different dependence structures.
Corollary~3.8 in \cite{MRu} provides for $\alpha \geq 1$ the inequalities
\begin{gather}\label{alphagreater1}
\limsup_{t\to \infty} \frac{\pr{ a_i V_{\ind}>t}}{\pr{ a_i V }}\leq 1\  \operatorname{and}\ \ \limsup_{t\to \infty} \frac{\pr{a_i V>t}}{\pr{ a_i  V_{\dep}>t }}\leq 1\ 
\end{gather}
and for $0<\alpha<1$ the inequalities 
\begin{gather*}\label{alphasmaller1}
\limsup_{t\to \infty} \frac{\pr{a_i V_{\dep}>t}}{\pr{ a_i V>t }}\leq 1\  \operatorname{and}\ \ \limsup_{t\to \infty} \frac{\pr{a_i V>t}}{\pr{ a_i V_{\ind}>t }} \leq 1.
\end{gather*}
Regarding the left inequality in \eqref{alphagreater1}, we have 
\begin{align} \nonumber
\limsup_{t \to \infty} \frac{\pr{ a_i  V_{\ind}>t}}{\pr{ a_i V >t }}
&=\limsup_{t \to \infty} 
\frac{\pr{ a_i  V_{\ind}>t}}{\pr{\|V_{\ind}\|>t}}\frac{\pr{\|V_{\ind}\|>t}}{\pr{V_{\ind,i}>t} }\Big/  \frac{\pr{ a_i  V>t}}{\pr{\|V\|>t}}\frac{\pr{\|V\|>t}}{\pr{{V}_i}>t}\\
&= \frac{\nu_{\ind}\circ A^{-1}(\{ x_i >t\})\nu(\{x_i>1 \})}{\nu \circ A^{-1}(\{ x_i>t\})\nu_{\ind}(\{x_i>1 \})}= 
\frac{\nu_{\ind} \circ A^{-1}(\{ x_i>1\})}{\nu \circ A^{-1}(\{ x_i>1\})}\leq 1\label{propuni},
\end{align} 
since w.l.o.g all marginals are the same.
The other inequalities in \eqref{Calphagreateroneindiv} as well as in  \eqref{Calphasmalleroneindiv} are treated analogously.
\end{proof}

For bounds on the market risk measures we invoke ideas from \cite{MRu}. 
{Below we sometimes write $C^{i}_{\nu}(A)$ and $C^{S}_{\nu}(A)$ instead of $C^{i}_{\nu}$ and $C^{S}_{\nu}$, if we want to emphasize that the constants depend on a particular matrix $A$.}

\bthe\label{thind}
Let the three $d$-dimensional vectors  $V_{\ind}$, $V$ and $V_{\dep}$ be given with equal margins $V_1,\dots,V_d$ with $\pr{V_j>t}\sim K_j t^{-\al}$, {but different exponent measures $\nu_{\ind}, \nu, \nu_{\dep}$}.
Denote the aggregated vector $\|F\|$ for some $r$-norm for $r \geq 1$ {or $r$-quasinorm for $0<r<1$}, representing the risk in the market. \\
(a) \, If $r\geq 1,$ for the constants  $C^S$ referring to the system setting risk the following inequalities hold:
\beam
\label{sysconstupperind}C^S_\nu  & \geq & C^S_{\ind} \quad \operatorname{for}\ \alpha\geq r,  \\
\label{sysconstlowerind}C^S_\nu   & \leq & C^S_{\ind} \quad \operatorname{for}\  0<\alpha \leq 1.
\eeam
{(b) \, If $0<r<1$, for the constants  $C^S$ referring to the  system setting the following inequalities hold:
\beam
\label{sysconstupperindquasi} C^S_\nu  & \geq & C^S_{\ind} \quad \operatorname{for}\ \alpha \geq 1 ,  \\
\label{sysconstlowerindquasi} C^S_\nu   & \leq & C^S_{\ind} \quad \operatorname{for}\  0<\alpha \leq  r.
\eeam
(c) \, However, there are matrices $A_{1}, A_{2}$ and an exponent measure $\nu_0$ such that
\beam\label{indcounter1}
C^{S}_{\ind}(A_1) & > & C^{S}_{\nu_0}(A_1)  \quad \operatorname{for}\ 1< \alpha <r ,\\\label{indcounter2} 
C^{S}_{\nu_0}(A_2)& > & C^{S}_{\ind}(A_2)   \quad \operatorname{for}\ 1<\alpha <r,\\\label{indcounter3} 
C^{S}_{\ind}(A_1) & < & C^{S}_{\nu_0}(A_1)  \quad \operatorname{for}\ r<\alpha < 1,\\\label{indcounte4} 
C^{S}_{\nu_0}(A_2)& < & C^{S}_{\ind}(A_2)   \quad \operatorname{for}\ r<\alpha < 1, 
\eeam}
\ethe

\begin{proof}
(a) \, In analogy to \cite{MRu} we define for $s^{1/\al}:=(s_1^{1/\al},\dots,s_d^{1/\al})$
\beao
g_{A,\al}(s) :=\|As^{1/\al}\|^{\al}\quad\mbox{and}\quad \rho^{*}g_{A,\al}:=\int_{\mathbb{S}_{+}^{d-1}}g_{A,\al}(s)\rho^{*}(ds)
\eeao
 for some canonical spectral measure $\rho^{*}$. 
 Similar to \eqref{propuni}, we note that
\begin{gather}
\frac{\nu_{\ind}\circ A^{-1}(\{ \| x\| >1\}) }{\nu\circ A^{-1}(\{ \|x\|>1\})} =\lim_{t\to \infty }\frac{\pr{\|AV_{\ind}\|>t}}{\pr{\|AV\|>t}}.
\end{gather}
Furthermore, we get from Propositions~3.2 and~3.3 in \cite{MRu} that 
\begin{gather}
\lim_{t\to \infty }\frac{\pr{\|AV_{\ind}\|>t}}{\pr{\|AV\|>t}}= \frac{ \rho^{*}_{\ind}g_{A,\al}}{ \rho^{*} g_{A,\al}}
\end{gather}
holds.
Hence, in order to prove \eqref{sysconstupperind} and \eqref{sysconstlowerind} it is sufficient to show that
$ \rho_{\ind}^{*}(g_{A,\al}) \leq \rho^{*}(g_{A,\al})  $ and $ \rho_{\ind}^{*}(g_{A,\al}) \geq \rho^{*}(g_{A,\al})  $, respectively. \\[2mm]
We first show \eqref{sysconstupperind}. 
Note that for nonnegative real numbers $a_1,\dots,a_n$ and $\beta \geq1$  the inequality 
\begin{gather}\label{simbinom}
a_1^{\beta}+ \dots + a_n^{\beta}\leq (a_1+\dots+a_n)^{\beta}
\end{gather}
is  valid.
Since $\rho_{\ind}^{*}g_{A,\al}= \sum_{j=1}^{d}\|Ae_j\|^{\al}$, and using \eqref{canspec}, we write as in the proof of Theorem~3.7 of \cite{MRu}
\begin{gather*}
\rho_{\ind}^{*}g_{A,\al}=\int_{\mathbb{S}_{+}^{d-1}}\sum_{j=1}^{d}\|Ae_j\|^{\al} s_j \rho^{*}(ds)=\int_{\mathbb{S}_{+}^{d-1}}\frac{\sum_{j=1}^{d}\|A s_j^{1/\al}e_j\|^{\al}}{\|\sum_{j=1}^{d} A s_j^{1/\al}e_j  \|^{\al}} \|As^{1/\al}\|^{\al} \rho^{*}(ds).
\end{gather*}
In order to establish $\rho_{\ind}^{*}g_{A,\al}\leq \rho^{*}g_{A,\al}$ it is sufficient to bound the fraction under the right hand integral by one. 
For this, we recall that all the entries in $A$ are nonnegative and that $\frac{\alpha}{r}\geq1$. 
We compute 
\begin{align}
\sum_{j=1}^{d}\|A s_j^{1/\al}e_j\|^{\al}
=\sum_{j=1}^{d} \Big( \sum_{i=1}^{q} (a_{ij}s_j^{1/\al})^r   \Big)^{\frac{\al}{r}} \label{wecomputefirst}
&\leq  \Big( \sum_{j=1}^{d}\sum_{i=1}^{q} (a_{ij}s_j^{1/\al})^r    \Big)^{\frac{\al}{r}} \\ \label{wecomputesecond}
&\leq  \Big(\sum_{i=1}^{q} \Big( \sum_{j=1}^{d}a_{ij}s_j^{1/\al}\Big)^r  \Big)^{\frac{\al}{r}} \\
& =\| \sum_{j=1}^{d}  A s_j^{1/\al}e_j \|^{\al} \nonumber
\end{align} 
where we have applied inequality \eqref{simbinom} twice.\\[2mm]
For the bound \eqref{sysconstlowerind}  we  use the $c_r-$inequality, see e.g. \cite{loeve}, p.~157, leading to
\begin{gather}\label{crineq}
\|\sum_{i=1}^{n} x_i\|^{\al}\leq \Big( \sum_{i=1}^{n} \|x_i\|  \Big)^{\al}\leq \sum_{i=1}^{n} \|x_i\|^{\al}
 \end{gather}
for $x_1,\dots,x_n \in \R^{d}.$
In particular, 
\beao
\rho^{*}_{\ind}g_{A,\al} &=& \sum_{j=1}^{d}\|Ae_j\|^{\al}= \int_{\mathbb{S}_{+}^{d-1}}\sum_{j=1}^{d} \|Ae_j\|^{\al} s_j d\rho^{*}(s)\\ &=& \int_{\mathbb{S}_{+}^{d-1}} g_{A,\al}(s)\frac{ \sum_{j=1}^{d} \|A s_{j}^{1/\alpha} e_j\|^{\al}}{\|As^{1/\al}\|^{\al}}  d\rho^{*}(s) 
\ \geq \ \rho^{*}g_{A,\al} 
\eeao
 leading to 
 $$C_\nu^S=\nu\circ A^{-1}(\{ \|x\|>1 \}) \leq \nu_{\ind}\circ A^{-1}(\{ \|x\|>1 \}) = C_{\ind}^S$$
 as expressed in \eqref{sysconstlowerind}.\\[2mm]
{(b) \,  We can proceed analogously to the proof of part (a), simply reversing inequalities.
 In order to establish \eqref{sysconstlowerindquasi}, we note that inequality \eqref{simbinom} holds in a reverse way for $\beta \leq 1.$
 Consequently, also both inequalities \eqref{wecomputefirst} and \eqref{wecomputesecond} hold analogously in the opposite way. 
For \eqref{sysconstupperindquasi}, note that in the case of $\alpha \geq 1$ and $0<r<1$ both inequalities in \eqref{crineq} obviously hold in a reverse way. 
\\[2mm]}
(c) \, Concerning examples for \eqref{indcounter1} and \eqref{indcounter2}, we choose $\nu_0$ to be the image measure $\nu_0:=\nu_{\ind}\circ B^{-1}$ with  standard exponent measure $\nu_{\ind}$ on $\R_+^{3}$ given as usual by $\nu_{\ind}([0,x]^c)= \sum_{j=1}^{3}x_j^{-\al}$ and a matrix 
$$B=\begin{pmatrix}
1 & 1 & 0\\
1 & 0 & 1
\end{pmatrix}.$$ 
Furthermore, we define the function $T:\R_{+}^{2}\rightarrow \R^{2}_{+}$  as 
$$T(x)=((\nu_{0}(\{y\in \R_+^{2}:|y_1|>1\})x_1)^{1/\al},(\nu_{0}(\{y\in \R_+^{2}:|y_2|>1\})x_2)^{1/\al})^{\top}.$$
The measure $\nu_0^{*}=\nu_{0}\circ T$ is then canonical; i.e., it is homogeneous of order $-1$   and  $\nu_0^{*}(\{y\in \R_+^{2}:|y_i|>1\})=1$ for $i=1,2 $.
To get the canonical spectral measure, we conduct the transformation to polar coordinates by setting $\tau(x)=(\|x\|,\frac{x}{\|x\|})$.
Denoting by $\rho^{*}_0$ the spectral measure and defining the measure $\pi$ by $d\pi(x)=x^{-2}dx$, the relation  $\nu_{0}^{*}= \pi \otimes \rho_1^{*}$ holds. We can now calculate $\rho_0^{*}$ as follows. 
We first note that by construction $\nu_0$ and hence $\nu_0^{*}$ only have positive mass on the axes as well as on the diagonal $\{ t \1: \ t>0\}$. Therefore, the canonical spectral measure, living on the sphere $\mathbb{S}_{+}^{d}$, only attains mass at the points $(1,0)^{\top}, (0,1)^{\top}, \1/\|\1\|$.
We first observe that $\nu_0 \circ B^{-1}(\{x:|x_i|>1\})=2$ for $i=1,2.$ 
This yields
\beao
\rho_0^{*}(\{(1,0)^{\top}\}) &=& \nu_0\circ T(\{ t e_1  | \ t>1    \}) \\
&=& \nu_{\ind}\circ B^{-1}(\{ 2^{1/\al} t e_j  | \ t>1    \})\\
&= &\nu_{\ind}(x\in \R_{+}^{3}| \ Bx \in \{ 2^{1/\al} t e_1 \in \R_+^{2}  | \ t>1    \} )\\
&=& \nu_{\ind}(s e_2 \in \R_{+}^{3}|\ sBe_2 \in \{ 2^{1/\al} t e_1 \in \R_+^{2}  | \ t>1    \} )\\ 
&= & \nu_{\ind}(s e_2 \in \R_{+}^{3}| \  s \in [2^{1/\al},\infty)   ) \ = \ \frac12 \ = \ \rho_0^{*}(\{(0,1)^{\top}\})
\eeao
by symmetry. 
For the third atom we calculate
\beao
 \rho_0^{*}(\{\1/\|\1\|\})&=& \nu_0\circ T(\{ t \1/\|1\|  | \ t>1    \})\\
 &= &\nu_{\ind}\circ B^{-1}(\{ 2^{1/\al} t \1/\|\1\|^{1/\al}  | \ t>1    \})\\
&= & \nu_{\ind}(x\in \R_{+}^{3}| \ Bx \in \{ 2^{1/\al} t \1/\|\1\|^{\1/\al} \in \R_+^{2}  | \ t>1    \} )\\
&=& \nu_{\ind}(s e_1 \in \R_{+}^{3}|sBe_1 \in \{ (2/\|1\|)^{1/\al} t \1 \in \R_+^{2}  | \ t>1    \} )\\ 
&= & \nu_{\ind}(s e_1 \in \R_{+}^{3}|\ s\in [(2/\|\1\|)^{1/\al},\infty)   )=\frac{\|\1\|}{2}.
\eeao
Consequently, we have
\begin{gather*}
\rho_{0}^{*}= \frac12 \delta_{(1,0)^{\top}} + \frac12 \delta_{(0,1)^{\top}} + \frac{\|\1\|}{2}\delta_{\1/\|\1\|}.
\end{gather*}
Furthermore, the canonical spectral measures for the case of asymptotical independence and full dependence are
\begin{gather*}
\rho_{\ind}^{*}= \delta_{(1,0)^{\top}} + \delta_{(0,1)^{\top}}\ \operatorname{and}\  \ \rho_{\dep}^{*}=\|\1\| \delta_{\1/\|\1\|}
\end{gather*} 
In order to construct counterexamples we choose $d=q=2$ and the function $g_{A_1,\alpha}$ with $A_1= I_2$ the identity matrix.
Then
\begin{align*}
\rho^{*}_{0}g_{A_1,\al}&= \int_{\mathbb{S}_+^1} \|A_1 s^{1/\al}\|^{\al}d\rho_{0}^{*} \\
&= \|I_2 (1,0)^{\top}\|^{\al} \rho_{1}^{*}(\{(1,0)^{\top}\})+\|I_2 (0,1)^{\top}\|^{\al}\rho_{1}^{*}(\{(0,1)^{\top}\})+\|I_2 (\1/\|\1\|)^{1/\al}\|^{\al} \rho_{1}^{*}(\{\1/\|\1\|\})\\
&= 2^{-1} +  2^{-1} + \|\1\|^{-1} \|(1,1)^{\top}\|^{\al}\frac{\|\1\|}{2}\\
& =1+ 2^{\frac{\al}{r}-1} ,
\end{align*}
while 
$\rho_{\ind}^{*}g_{A_1,\al}=2$. 
This leads to  the equivalences  
\begin{align}\label{firstcounter}
\rho_{0}^{*}g_{A_1,\al}<\rho_{\ind}g_{A_1,\al} \ \Leftrightarrow \ 2>1+ 2^{\frac{\al}{r} - 1}  \ \Leftrightarrow 	\ 1 >2^{\frac{\al}{r} - 1}  \ \Leftrightarrow \ r >\al .
\end{align}
 In particular, we have  for $1< \al < r,$ 
\begin{gather*}
C^{S}_{\nu_0}(A_1) < C^{S}_{\ind}(A_1).
\end{gather*}
Next, we choose $A_2= \begin{pmatrix} 1 &1 \\ 1 & 1 \end{pmatrix}$ and calculate 
$$\rho^{*}_{\ind}g_{A_2,\al}= \left\|\1\right\|^{\al}  + \left\| \1 \right\|^{\al} =  2^{\frac{\al}{r}+1}$$
as well as 
\begin{align*}
\rho^{*}_{0}g_{A_2,\al} = \frac12 \left\|\1 \right\|^{\al} +\frac12 \left\| \1 \right\|^{\al} + \frac{\| \mathds{1}\|}{2} \left\| \begin{pmatrix} 1&1\\ 1 & 1 \end{pmatrix} \frac{\mathds{1} }{\|\mathds{1}\|^{1/\alpha}}  \right \|^{\al}.
\end{align*}
Consequently,
\begin{gather}\label{secondcounter}
  \rho^{*}_{\ind}g_{A_2,\al}<  \rho^{*}_{0}g_{A_2,\al} \ \Leftrightarrow \ 2< 2^\al 
	\end{gather}
Therefore, for $\alpha >1,$ $C^S_{\ind}(A_2) <  C^{S}_{\nu_0}(A_2).$
{Inequalities \eqref{indcounter3} and \eqref{indcounte4} follow then from \eqref{firstcounter} and \eqref{secondcounter}, respectively.} 
\end{proof}

\bthe\label{thdep}
Let the three $d$-dimensional vectors  $V_{\ind}$, $V$ and $V_{\dep}$ be given with equal margins $V_1,\dots,V_d$ with $\pr{V_j>t}\sim K_j t^{-\al}$, {but different exponent measures $\nu_{\ind}, \nu, \nu_{\dep}$}.
Denote the aggregated vector $\|F\|$ for some $r$-norm for $r>1$ {or some $r$-quasinorm for $0<r<1$}, representing the risk in the market.\\
(a)\, If $r\geq 1,$ for the constants $C^S$ referring to the system setting the following inequalities hold:
\beam
\label{sysconstupperdep}C^S_\nu  &\leq &  C^S_{\dep} \quad \operatorname{for}\ \alpha\geq r  \\
\label{sysconstlowerdep}C^S_\nu  & \geq & C^S_{\dep} \quad \operatorname{for}\  0<\alpha \leq  1
\eeam
{(b)\, If $0< r <1,$ for the constants $C^S$ referring to the system setting the following inequalities hold:
\beam
\label{sysconstupperdepquasi}C^S_\nu  &\leq &  C^S_{\dep} \quad \operatorname{for}\ \alpha \geq 1  \\
\label{sysconstlowerdepquasi}C^S_\nu  & \geq & C^S_{\dep} \quad \operatorname{for}\  0<\alpha \leq  r
\eeam
(c)\, However,  there are matrices $A_{1}, A_{2}$ and an  exponent measure $\nu_0$ such that
\beam
\label{counterdep1} C^{S}_{\nu_0}(A_1) &>& C^{S}_{\dep}(A_1)  \quad \operatorname{for}\ 1< \alpha <r ,\\
\label{counterdep2}C^{S}_{\dep}(A_2) &>& C^{S}_{\nu_0}(A_2) \quad \operatorname{for}\ 1<\alpha <r ,\\
\label{counterdep3} C^{S}_{\nu_0}(A_1) &<& C^{S}_{\dep}(A_1)  \quad \operatorname{for}\ r< \alpha <1 ,\\
\label{counterdep4}C^{S}_{\dep}(A_2) &<& C^{S}_{\nu_0}(A_2) \quad \operatorname{for}\ r<\alpha <1 .
\eeam
}
\ethe

\begin{proof}
We need the following inequalities, which are known as generalizations of Theorem~202 in \cite{hardy}, where such inequalities are proved for integrals with respect to Lebesgue measures. The general versions below are natural extensions using Fubini's theorem and the H\"older inequality for $\sigma$-finite measures.
Suppose $(S_1,\mu_1), (S_2,\mu_2)$ are two $\sigma$-finite  measure spaces and $F:S_1\times S_2\rightarrow \R$ is a product-measurable mapping. Then for $p \geq 1$ the inequality 
\begin{gather} \label{intmink}
\int_{S_2}\left|     \int_{S_1} F(x,y) d\mu_1(x)  \right|^{p} d\mu_2(y) \leq \left( \int_{S_1} \left(  \int_{S_2} |F(x,y) |^{p} d\mu_2(x)\right)^{\frac{1}{p}} d\mu_1 \right)^{p}
\end{gather} 
and for $0<p\leq 1 $ the inequality 
\begin{gather} \label{intminkrev}
\int_{S_2}\left|     \int_{S_1} F(x,y) d\mu_1(x)  \right|^{p} d\mu_2(y) \geq \left( \int_{S_1} \left(  \int_{S_2} |F(x,y) |^{p} d\mu_2(x)\right)^{\frac{1}{p}} d\mu_1 \right)^{p}
\end{gather} 
hold true. \\[2mm]
(a) \,  In the case $1<r<\alpha$ we want to show \eqref{sysconstupperdep}; more precisely,
\begin{gather} \label{depupper}
\int_{\mathbb{S}_{+}^{d-1}} \|As^{1/\alpha}\|^{\al}d\rho^{*}(s)\leq  \int_{\mathbb{S}_{+}^{d-1}} \|As^{1/\alpha}\|^{\al}d\rho_{\dep}^{*}(s)  = \|A\1\|^{\al}.
\end{gather}
To this end, we will apply \eqref{intmink} twice. In a first step, take
$S_{2}=\mathbb{S}_{+}^{d-1}$ with $\mu_2=\rho$ and $S_{1}=\{1,\dots,q \}$ with $\mu_1$ the counting measure, as well as
$F(i, s)= \big(\sum_{j=1}^{d} A_{ij}s_{j}^{1/\al} \big)^{r}$ and $p=\frac{\al}{r}.$ Then  
\begin{align} \nonumber
\int_{\mathbb{S}_{+}^{d-1}} \|As^{1/\alpha}\|^{\al}d\rho^{*}(s)
&= \int_{S_2}\left|  \int_{S_1} F(x,y) d\mu_1(x)  \right|^{p} d\mu_2(y)\\ \nonumber
& \leq \Big( \int_{S_1} \Big(  \int_{S_2} |F(x,y) |^{p} d\mu_2(x)\Big)^{\frac1p} d\mu_1 \Big)^{p}\\ \label{continue}
&= \Big(  \sum_{i=1}^{q} \Big(\int_{\mathbb{S}_{+}^{d-1}}  \Big(\sum_{j=1}^{d} A_{ij}s_{j}^{1/\al}  \Big)^{r\frac{\al}{r}} d\rho^{*}(s)\Big)^{\frac{r}{\al}} \Big)^{\frac{\al}{r}}
 \end{align}
In the second step, take $S_2=\mathbb{S}_{+}^{d-1}$ with $\mu_2=\rho^{*}$ and $S_{1}=\{1,\dots,d\}$ with the  weighted counting measure $\mu^{i}_1=\sum_{j=1}^{d}A_{ij}\delta_{j} $ for $i=1,\dots,q$. Further, let $F(j,s)=s_{j}^{1/\al}$ and $p=\al.$ 
Then  
\begin{align*}
\int_{\mathbb{S}_{+}^{d-1}}  \Big(\sum_{j=1}^{d} A_{ij}s_{j}^{1/\al}  \Big)^{\al} d\rho^{*}(s)
&\leq \Big( \sum_{j=1}^{d} A_{ij}  \Big(\int_{\mathbb{S}_{+}^{d-1}} (s_{j}^{1/\al})^{\al} d\rho^{*}(s) \Big)^{1/\al}  \Big)^{\al} = (\sum_{j=1}^{d}A_{ij})^{\al},\ i=1,\dots, q. 
\end{align*}
We continue with \eqref{continue} and find that
\begin{align*}
\Big(  \sum_{i=1}^{q} \Big(\int_{\mathbb{S}_{+}^{d-1}}  \Big(\sum_{j=1}^{d} A_{ij}s_{j}^{1/\al}  \Big)^{r\frac{\al}{r}} d\rho^{*}(s)\Big)^{\frac{r}{\al}} \Big)^{\frac{\al}{r}}
\leq \Big(  \sum_{i=1}^{q} \Big((\sum_{j=1}^{d}A_{ij})^{\al}\Big)^{\frac{r}{\al}} \Big)^{\frac{\al}{r}}= \|A\1\|^{\al}.
\end{align*}
Relation \eqref{sysconstlowerdep} is shown analogously using \eqref{intminkrev}.\\[2mm]
{(b) \, Inequalities \eqref{sysconstupperdepquasi} and \eqref{sysconstlowerdepquasi}  can be shown similar to part (a) by using the respective reverse inequalities.
}\\[2mm](c) \, Finally, we can use $\rho_{0}^{*}$ in order to show \eqref{counterdep1} and \eqref{counterdep2}. 
 Taking again $A_1=I_2$, we obtain
$$
\rho_{0}^{*}g_{A_1,\al}= 1+ 2^{\frac{\al}{r} -1}\quad\mbox{and}\quad
 \rho_{\dep}^{*}g_{A_1,\al}=2^{\frac{\al}{r}}$$ 
and, consequently, 
\begin{gather}\label{secondfirstcounter}
\rho_{0}^{*}g_{A_1,\al} \ > \ \rho_{\dep}^{*}g_{A_1,\al}  \ \Leftrightarrow \ 1+ 2^{\frac{\al}{r} -1} > 2^{\frac{\al}{r}} \ \Leftrightarrow \ 2 > 2^{\frac{\al}{r}}\Leftrightarrow  \alpha <r .\end{gather}
Therefore,  we have $ C_{\nu_0}^{S}(A_1)> C^{S}_{\dep}(A_1)$  for $1< \al <r$.  \\
Next, we choose $A_2:=\begin{pmatrix} 1 & 1 \\ 1 & 1 \end{pmatrix}$ and compute
$$\rho_{1}^{*}g_{A_2,\al}= 2^{\frac{\al}{r}} + 2^{-1}2^{\al(1+\frac{1}{r})} \quad\mbox{and}\quad
\rho_{\dep}^{*}g_{A_2,\al}=2^{\al(1+\frac{1}{r})}.$$
As a matter of fact, 
\begin{align}\label{secondsecondcounter}
\rho_{\dep}^{*}g_{A_2,\al}>\rho_{1}^{*}g_{A_2,\al}  \ \Leftrightarrow \ 2^{\al}> 2 \Leftrightarrow \alpha >1;
\end{align}
i.e., we have $C_{\nu_0}^{S}(A_2) < C_{\dep}^{S}$ for $1<\al <r$. {Relations  \eqref{counterdep3} and \eqref{counterdep4} then follow from \eqref{secondfirstcounter} and \eqref{secondsecondcounter}, respectively.} 
\end{proof}

\begin{corollary}Given the assumptions of Theorems~\ref{thind} and \ref{thdep}, respectively,  in the particular case
$\alpha=r=1$ the equalities
\begin{gather*}
C_{\ind}^{S}=C_{\dep}^{S}=C_{\nu}^{S}
\end{gather*}hold true; i.e., the Value-at-Risk asymptotics are not influenced by the dependence structure between the risk factors. 
\end{corollary}

\brem 
Considering asymptotic bounds for the $\VaR_{1-\gamma}$ and the $\CoTE_{1-\gamma}$ as in this paper can be of practical relevance since $\gamma$ is typically taken to be small.
Bounds for tail risk measures have also been studied in \cite{BRV} in a non-asymptotic setting. From these bounds, one can derive asymptotic versions. 
As an example consider the upper bounds, while setting $q=d$ and $A=Id$ (the identity matrix, which is in particular deterministic) as well as $\alpha \in (1,\infty)$ and $\|\cdot\|_1$ as the norm. 
Then Theorem \ref{thdep} implies for an arbitrary risk vector $V$  
\begin{gather}\label{upper-dep-inf-var} 
\VaR_{1-\gamma}(\|V\|_1) \preceq \VaR_{1-\gamma}(\|V_{\dep}\|_1)\end{gather}
(where $f\preceq g $ is defined as $\limsup_{\gamma\to 0} \frac{f(1-\gamma)}{g(1-\gamma)}\leq 1$ ).
If the distribution of $\|V\|_1$ has a density, then
the Conditional Tail Expectation $\CoTE_{1-\gamma}(\|V\|_1)$ from Definition~\ref{defVaRCoTE}  coincides with what is introduced as  
\textit{Tail Value-at-Risk} $\operatorname{TVaR_{1-\gamma}}(\|V\|_1)$ in \citep{BRV}.\\[2mm]
If $\|V\|_1$ has infinite variance corresponding to $ \alpha \in (1,2]$, we get from Theorem 2.1 in \cite{BRV}  the inequality
 $\VaR_{1-\gamma}(\| V \|_1) \leq \CoTE_{1-\gamma}(\|V\|_1),$
   which obviously implies also 
   $$\VaR_{1-\gamma}(\| V \|_1) \preceq \CoTE_{1-\gamma}(\|V_{\dep} \|_1).$$ 
  If we combine this with  the Karamata asymptotic
  $$\CoTE_{1-\gamma}(\|V_{\dep}\|_1)\sim \frac{\alpha}{\alpha-1} \VaR_{1-\gamma}(\|V_{\dep}\|), $$
this gives
$$\VaR_{1-\gamma}(\|V\|_1) \preceq  \frac{\alpha}{\alpha-1} \VaR_{1-\gamma}(\|V_{\dep}\|_1),$$ 
which is in the light of \eqref{upper-dep-inf-var} not an optimal bound as $\frac{\alpha}{\alpha-1}>1$ for $\alpha >1$.\\[2mm]
 If $\|V\|_1$ has finite variance, corresponding to $\alpha>2$, Theorem 3.2 of \citep{BRV} gives the inequality 
 $$ \VaR_{1-\gamma}(\|V\|_1)\leq \min\Big\{ \mu + s\sqrt {(1-\gamma)/{\gamma}},\CoTE_{1-\gamma}(\|V_{\dep}\|_1) \Big\}$$ 
 with $\mu=\E \|V\|_1$ and $s^2=\var \|V\|_1 $. 
 Since $\gamma^{-1/\alpha}\ \preceq \sqrt{{(1-\gamma})/{\gamma}}$ as $\gamma \to 0,$ this  also is no optimal bound compared to \eqref{upper-dep-inf-var}.
\erem

\section*{Acknowledgements}
We are grateful to Ludger R\"uschendorf for making us aware of results in the paper \cite{MRu}.

\begin{small}
\bibliography{bibgesinebounds}

\begin{thebibliography}{10}

\bibitem{Basrakarticle}
B.~Basrak, R.A. Davis, and T.~Mikosch.
\newblock Regular variation of {GARCH} processes.
\newblock {\em Stochastic Processes and their Applications}, 99(1):95 -- 115,
  2002.

\bibitem{Beirlant}
J.~Beirlant, Y.~Goegebeur, J.~Segers, and J.~Teugels.
\newblock {\em {Statistics of Extremes: Theory and Applications}}.
\newblock Wiley Series in Probability and Statistics. Wiley, Chichester, 2006.

\bibitem{BRV}
C.~Bernard, L.~R\"uschendorf, and S.~Vanduffel.
\newblock Value-at-risk bounds with variance constraints.
\newblock Forthcoming in \textit{Journal of Risk and Insurance}. Available at
  SSRN: \url{http://ssrn.com/abstract=2342068}, 2016.

\bibitem{OpRiskBK}
K.~B\"ocker and C.~Kl\"uppelberg.
\newblock Multivariate models for {O}perational {R}isk.
\newblock {\em Quantitative Finance}, 10(8):855--869, 2010.

\bibitem{Breiman}
L.~Breiman.
\newblock {On some limit theorems similar to the arc-sine law}.
\newblock {\em Theory Probab. Appl.}, 10:323--331, 1965.

\bibitem{BurgertRü}
C.~Burgert and L.~R\"uschendorf.
\newblock {Consistent risk measures for portfolio vectors}.
\newblock {\em Insurance: Mathematics and Economics}, 38(2):289--297, 2006.

\bibitem{Cacciolietal}
F.~Caccioli, M.~Shrestha, C.~Moore, and J.D. Farmer.
\newblock Stability analysis of financial contagion due to overlapping
  portfolios.
\newblock {\em Journal of Banking \& Finance}, 46:233 -- 245, 2014.

\bibitem{axiomsystemic}
C.~Chen, G.~Iyengar, and C.C. Moallemi.
\newblock An axiomatic approach to systemic risk.
\newblock {\em Management Science}, 59(6):1373--1388, 2013.

\bibitem{Embrechts}
P.~Embrechts, D.D. Lambrigger, and M.V. W\"uthrich.
\newblock Multivariate extremes and the aggregation of dependent risks:
  examples and counter-examples.
\newblock {\em Extremes}, 12(2):107--127, 2009.

\bibitem{EPR}
P.~Embrechts, G.~Puccetti, and L.~R\"uschendorf.
\newblock Model uncertainty and {V}a{R} aggregation.
\newblock {\em Journal of Banking and Finance}, 37(8):2750--2764.

\bibitem{fernholz}
R.~Fernholz, R.~Garvy, and J.~Hannon.
\newblock {Consistent risk measures for portfolio vectors}.
\newblock {\em Journal of Portfolio Management}, 24(2):74--82, 1998.

\bibitem{hardy}
G.H. Hardy, J.E. Littlewood, and G.~P{\'o}lya.
\newblock {\em Inequalities}.
\newblock Cambridge Mathematical Library. Cambridge University Press, 1952.

\bibitem{KKR2}
O.~Kley, C.~Kl\"uppelberg, and G.~Reinert.
\newblock Conditional risk measures in a bipartite market structure.
\newblock Submitted, 2015.

\bibitem{KKR1}
O.~Kley, C.~Kl\"uppelberg, and G.~Reinert.
\newblock Risk in a large claims insurance market with bipartite graph
  structure.
\newblock Forthcoming in \textit{Operations Research}, 2016.

\bibitem{loeve}
M.~Loeve.
\newblock {\em Probability Theory, Vol.~1}.
\newblock Springer, New York, 4 edition, 1977.

\bibitem{MRu}
G.~Mainik and L.~R\"uschendorf.
\newblock Ordering of multivariate risk models with respect to extreme
  portfolio losses.
\newblock {\em Statistics \& Risk Modeling}, 29(1):73--106, 2012.

\bibitem{Resnick2007}
S.I. Resnick.
\newblock {\em Heavy-Tail Phenomena}.
\newblock Springer, New York, 2007.

\end{thebibliography}
\bibliographystyle{plain}
\end{small}

\end{document}